\documentclass[floatfix,aps,prd,amsmath,nofootinbib,onecolumn,11pt]{revtex4-1}
\def \be {\begin{equation}} 
\def \ee {\end{equation}} 
\def \bea {\begin{eqnarray}} 
\def \eea {\end{eqnarray}} 

\usepackage{graphicx}
\usepackage{dcolumn}
\usepackage{bm}
\usepackage{epsfig} 
\usepackage{amsfonts}
\usepackage{amsmath}
\usepackage{amssymb}
\usepackage[usenames]{color}
\usepackage[dvipsnames]{xcolor}
\usepackage[unicode, colorlinks=true, linkcolor=linkcolor, citecolor=linkcolor, filecolor=linkcolor,urlcolor=linkcolor, pdfusetitle]{hyperref}

\hypersetup{colorlinks,citecolor=blue,linkcolor=blue,urlcolor=blue}
\hypersetup{final=true}

\begin{document}

\title{Measuring the speed of light with cosmological observations: current constraints and forecasts}

\author{Jaiane Santos}
\email{jaianesantos@on.br}
\affiliation{Observat\'orio Nacional, 20921-400, Rio de Janeiro - RJ, Brazil}

\author{Carlos Bengaly}
\email{carlosbengaly@on.br}
\affiliation{Observat\'orio Nacional, 20921-400, Rio de Janeiro - RJ, Brazil}

\author{Jonathan Morais}
\email{jonathanmorais@ufrrj.br}
\affiliation{Observat\'orio Nacional, 20921-400, Rio de Janeiro - RJ, Brazil}
\affiliation{Departamento de Física, Universidade Federal Rural do Rio de Janeiro, 23897-000, Seropédica - RJ, Brazil}

\author{Rodrigo S. Gonçalves}
\email{rsg\underline{ }goncalves@ufrrj.br}
\affiliation{Observat\'orio Nacional, 20921-400, Rio de Janeiro - RJ, Brazil}
\affiliation{Departamento de Física, Universidade Federal Rural do Rio de Janeiro, 23897-000, Seropédica - RJ, Brazil}

\date{\today}

\begin{abstract}

We measure the speed of light with current observations, such as Type Ia Supernova, galaxy ages, radial BAO mode, as well as simulations of forthcoming redshift surveys and gravitational waves as standard sirens. By means of a Gaussian Process reconstruction, we find that the precision of such measurements can be improved from roughly 6\% and to about $2-2.5\%$ when the gravitational wave simulations are considered, and to $1.5-2\%$ when redshift survey are included in the analysis as well. This result demonstrates that we will be able to perform a cosmological measurement of a fundamental physical constant with significantly improved precision, which will help us underpinning if its value is truly consistent with local measurements, as predicted by the standard model of Cosmology.  

\end{abstract}


\maketitle

\section{Introduction}\label{sec:intro}

Since the late 1990s, the Standard Cosmological Model (SCM) has been described by the flat $\Lambda$CDM model \cite{Riess,Perlmutter}, which states that the Universe is dominated by cold dark matter as the responsible for structure formation and galaxy dynamics, and by the cosmological constant $\Lambda$ as the responsible for the accelerated expansion of the Universe at late times. The framework of the SCM provides a series of successful predictions, such as recent observations of the Cosmic Microwave Background (CMB) \cite{Aghanim}, luminosity distances of Type Ia Supernovae (SNe) \cite{SCOLNIC}, as well as galaxy clustering and weak lensing \cite{Alam,Asgari,Abbott,Secco}, which validate the SCM as the model that best describes the observed data with great precision.

However, we can mention some unresolved problems in relation to the SCM, such as problems of primordial singularity and cosmic coincidence. We also have tensions in measurements of some cosmological parameters, e.g. the tension of $\sim 5\sigma$ between the Hubble Constant $H_0$ measured in the late- and early-time Universe with SNe and CMB, respectively, being one of most evident at the current moment~\cite{DiValentino} -- see also~\cite{Perivolaropoulos} for a broad review on cosmological tensions. Recently, the first data release of the DESI telescope showed that there could be an evolution of the dark energy equation of state, thus hinting at a possible breakdown of the cosmological constant paradigm~\cite{DESI}. In such a scenario, it is necessary to propose and test alternative models, besides revisiting the SCM fundamentals, since further evidence for departures of the SCM predictions would suggest new physics at play, and require a complete reformulation of its panorama. The validity of two fundamental pillars of the SCM, namely the Cosmological Principle and General Relativity as a theory of gravity on large scales, have been confirmed in recent works, although some results may affirm the opposite~\cite{Perivolaropoulos}.

A possible route to propose alternative models to the SCM consists on studying the variability of the fundamental constants of nature, as discussed in~\cite{Dirac,Uzan1,Uzan2,Martins}. Experiments with this purpose have been carried out for centuries on Earth and in the Solar System to measure the values of fundamental constants, obtaining zero results for their variations and with supreme precision in their measurements. However, cosmological tests of the consistency of fundamental constants are very scarce and less precise, due to the difficulty in obtaining cosmological data at high redshifts that would make their achievements possible. At this point, it is crucial to take care when leveraging such models, as they can trigger additional problems in the physical laws under which these physical constants were constructed. For example, models that assume a variable speed of light (VSL) must reproduce the success of the Special Theory of Relativity in explaining at least thermodynamics and electromagnetism. Some of these models meet these requirements and can provide viable solutions to SCM problems~\cite{Moffat1,Barrow1,Albrecht,Barrow2,Barrow3,Clayton1,Avelino,Clayton2,Bassett,Magueijo1,Clayton3,Magueijo2,Ellis1,Ellis2,Magueijo3,Cruz1,Moffat2,Franzmann,Cruz2,Costa,Gupta,Lee1}.

Driven by this reason, some recent studies searched for possible evidence of VSL model, and carried out speed of light measurements with cosmological observations, mostly obtaining null evidence for the former, and results in good concordance with the measurements in local laboratories for the latter~\cite{Balcerzak1,Zhang1,Balcerzak2,Zhang2,SALZANO,VSalzano,Cai,Balcerzak3,SCao1,Dabrowski,Salzano1,Guedes,ZouDeng,Huerta,SCao2,Wang,AAlbert,Pan,Mendonca,GC,Lee6,Mukherjee1,Liu,Lee7,Cuzinatto,Zhang}. For instance, in~\cite{GC}, the authors used data from the Pantheon SNe compilation, besides measurements of the Hubble parameter obtained through the differential ages galaxies and the radial mode of baryon acoustic oscillations (BAO), to measure the speed of light by means of the methodology proposed by~\cite{VSalzano}. These analyses were done using Gaussian Processes, i.e., a non-parametric reconstruction method -- thus, no cosmological model is assumed {\it a prori} -- resulting in measurements of $\simeq 5\%$ precision at $z = 1.4-1.6$. Although these results agree within $2\sigma$ confidence level with typical speed of light measurements here on Earth, we should note that the availability of cosmological data at such redshift range is quite limited. Hence, the precision of those tests are inevitably affected by this fact, not to mention that some of the cosmological data could be biased towards the SCM scenario.

Hence, considering the caveats of the SCM, as previously discussed, and considering the importance of performing cosmological measurements of the fundamental constants in this context, we forecast the precision of speed of light measurements that can be reached by future cosmological observations. We produce simulations of the Hubble parameter measurements expected from upcoming redshift surveys, as well as luminosity distance measurements that are expected from future observations of gravitational wave (GW) events, by means of the standard siren method. Although the speed of the GW event GW170817 was precisely measured as being the same as the speed of light, which confirms the predictions of general relativity theory~\cite{Flanagan}, here we pursue a different approach to measure the speed of light from GW, as carried out in~\cite{VSalzano, GC}, given the advent of the standard siren measurements that upcoming GW experiments shall provide. Our main goal is to assess the precision that those speed of light measurements can be improved, when compared to current constraints.

The paper is organized as follows: in section 2, we describe the theoretical framework, followed by section 3, in which we present the data and simulations, section 4 presents our results. Finally, section 5 is dedicated to discussion and concluding observations.
\section{Theoretical Framework}\label{sec:theory}

In order to obtain a cosmological measurement of the speed of light, we adopt the method proposed by \cite{SALZANO}. This method consists of assuming a flat Universe described by the Friedmann-Lema\^itre-Robertson-Walker (FLRW) metric, so that the angular diameter distance is given by by the equation 
\begin{equation}
\label{eq:DAz}
    D_{A}(z)=\frac{1}{(1+z)}\int_{0}^{z}\frac{cdz}{H(z)} \,,
\end{equation}
where $c=c(z)$ is the speed of light -- here, written as a function of redshift for the sake of convenience -- and $H(z)$ is the Hubble parameter that gives the expansion rate of the Universe at a given redshift $z$.

Thus, deriving Eq.~\eqref{eq:DAz} with respect to redshift, we have
\begin{equation}\label{eq:deriv_Daz}
   \frac{\partial}{\partial z}[(1+z)D_{A}(z)]=\frac{c(z)}{H(z)} \,,
\end{equation}
so we can write $c(z)$ as
\begin{equation}\label{eq:cz}
    c(z)=H(z)[(1+z)D'_{A}(z)+D_{A}(z)] \,,
\end{equation}
where $D'_{A}(z)$ denotes the first derivative of the angular diameter distance in relation to redshift. Also, we can obtain the uncertainties through an error propagation, given as follows\footnote{Note that there is a typo in the expression of $\sigma^{2}_{c(z)}$ presented in~\cite{GC}, which has been fixed now.},
\begin{equation}\label{eq:sigma_cz}
   \begin{aligned}
        \sigma^{2}_{c(z)}= H'(z)[(1+z)D_A'(z)+D_A(z)]\sigma^{2}_{H(z)}+H(z)[(1+z)D_A''(z)+D_A(z)]\sigma^{2}_{D_A'(z)}+ \\ 
        + H(z)[(1+z)D_A'(z)]\sigma^{2}_{D_{A(z)}} \,.
    \end{aligned}
\end{equation}

In the SCM scenario, $D_{A}(z)$ is expected to reach its maximum value at around $z \sim 1.4 - 1.6$, depending on the cosmological parameters of the model, for instance, the matter density parameter and the Hubble Constant. Therefore, when we take the first derivative of the angular diameter distance at this maximum redshift ($z_m$), its value will be zero (i.e. $D_A'(z_m) = 0$). So, at such redshift, Eq.~\eqref{eq:cz} can be rewritten as
\begin{equation}\label{eq:czm}
    c(z_{m}) = D_{A}(z_{m})H(z_{m}) \,,
\end{equation}
once more, we can obtain the uncertainties through an error propagation given as follows,
\begin{equation}\label{eq:sigma_czm}
  \sigma^{2}_{c(z_{m})}= \left(H(z) \sigma_{D_{A}(z)}\right)^{2}+\left(D_{A}(z) \sigma_{H(z)}\right)^{2} \,.  
\end{equation}

Therefore, we can arrive at an expression for the speed of light at such \textbf{a} redshift that depends only on the angular diameter distance and the Hubble parameter, allowing us to measure \textbf{it} with higher precision because it does not depend on their respective derivatives, which suffer from smearing effect -- and thus much larger uncertainties. It is worthful to notice that because there is a degeneracy between a variable speed of light and cosmic curvature, we note that this method is only valid for flat FLRW models \cite{VSalzano,Dabrowski}. Still, recent measurements of the curvature of the Universe are in excellent agreement with a flat Universe, so we can safely make this assumption~\cite{Aghanim,SCOLNIC,Alam,Asgari,Abbott,Secco,DESI}. 

\section{Data and simulations}\label{sec:data_sim}

We use 1701 luminosity distance measurements of Type Ia Supernovae (SNe) from the Pantheon+ and SH0ES compilation, and 48 Hubble parameter measurements $H(z)$ obtained through differential galaxy ages and the radial mode of baryon acoustic oscillations (BAO)~\cite{MAGANA,MORESCO}, as our current cosmological data-sets\footnote{Note that the previous Pantheon compilation~\cite{SCOLNIC} was used in~\cite{GC}.}. In addition, we produce simulations of future cosmological observations, such as 1000 luminosity distance measurements from gravitational wave events (GW) as standard sirens, as expected by experiments such as LIGO~\cite{LIGO} and Einstein Telescope (ET)~\cite{ET}, along with 30 $H(z)$ measurements that are expected for the new generation of redshift surveys, as the case of J-PAS~\cite{J-PAS1}, detailed as follow:

\subsection{Type Ia Supernovae}

The latest SN compilation, namely the Pantheon+SH0ES data-set~\cite{PantheonPlus1} (see also~\cite{PantheonPlus2,SH0ES}), provides 1701 light curve measurements of 1550 SN objects in the redshift interval $0.001 < z < 2.26$. Hence, we have 1701 measurements of SN apparent magnitudes, $m_{B}$, which can be combined with the determination of the SN absolute magnitude given by~\cite{SCOLNIC} 
\begin{equation}\label{eq:MB}
    M_{B} = -19.25 \pm 0.03 \,.
\end{equation}
We can obtain the luminosity distances through
\begin{equation}\label{eq:DLz}
    D_{L}(z)=10^{\frac{m_{B}(z)-M_{B}-25}{5}} \,.
\end{equation}
Then we can convert the luminosity distances into angular diameter distances by means of the cosmic distance duality relation (CDDR), which reads
\begin{equation}\label{eq:CDDR}
    D_{A}(z)=D_{L}(z)(1+z)^{-2} \,.
\end{equation}

This relationship is valid for all models based on Riemannian geometry, being independent of Einstein's field equations, the FLRW metric, or the nature of dark matter and dark energy. The violation of the CDDR would only occur in the case of geometry non-Riemannian, the presence of a source of cosmic opacity in the Universe, or variations in fundamental physics, such as the Equivalence Principle and the fine structure constant. Nonetheless, recent works showed that the CDDR is validated through a variety of cosmological observations and approaches, as in~\cite{Goncalves,Mukherjee2,Bora,Renzi,Tonghua}.

\subsection{Gravitational Waves}

In order to obtain the simulated data sets of gravitational waves, we assume that the sources of the events are mergers of Neutron Stars-Black Hole (NS-BH) or Neutron Stars-Neutron Stars (NS-NS) binaries. They obey a redshift distribution given by
\begin{equation}
    P(z) \propto \frac{4\pi D_{c}^{2}R(z)}{H(z)(1+z)},
\end{equation}
where $d_c$ is the comoving distance and $H(z)$ is the Hubble parameter \cite{WZhao}. Both quantities are defined from a fiducial cosmology by assuming a flat $\Lambda$CDM model with $H^{\rm fid}_0 = 73.30 \pm 1.04 \, \mathrm{km \, s}^{-1} \, \mathrm{Mpc}^{-1}$, $\Omega^{\rm fid}_{\rm m} = 0.334 \pm 0.018$ and $\Omega^{\rm fid}_{\Lambda} = 1-\Omega_{\rm m}$, consistent with the Pantheon+SH0ES best-fit~\cite{PantheonPlus1}.

The factor $R(z)$ describes the evolution of the star formation rate \cite{CCutler} and it has a specific functional form for each gravitational wave source given by
\begin{equation}
R(z) = 
\begin{cases}
    1 + 2z \text{,} & z \leq 1 \\
    \frac{3}{4}(5-z) \text{,} & 1 < z < 5 \\
    0 \text{,} & z \geq 5
\end{cases}
\end{equation}

From the probability distribution for each gravitational source we randomly pick 1000 points in the redshift range $0 < z < 2.36$. We then obtain the luminosity distance ($D_L$) at those redshift with the fiducial model, and we perform a Monte Carlo simulation by assuming a gaussian distribution centered on these fiducial $D_L$. The standard deviation ($\sigma_{D_L}$) is related with an instrumental error ($\sigma_{D_L^{inst}}$) that can be obtained via a Fisher Matrix analysis and combined with an additional error from weak lensing ($\sigma_{D_L^{WL}}$) \cite{JFZhang},
\begin{equation}
\sigma_{D_L} = \sqrt{ \left( \sigma_{D_L^{inst}} \right)^2 + \left( \sigma_{D_L^{WL}} \right)^2 } = \sqrt{\left( \frac{2D_L}{\rho}\right)^{2}+ \left( 0.05zD_L\right)^2}.
\end{equation}

The parameter $\rho$ represents the signal-to-noise ratio of the detection (SNR) and is directly related to the amplitude of the gravitational wave ($A$), the interferometer antenna pattern ($F$), and the power spectrum density ($PSD$) \cite{JFZhang}.

We simulate the data from two interferometer configurations, with its specific $F$ and $PSD$. The first configuration is defined by assuming a perpendicular interferometer (as the LIGO collaboration), in this case we have the antenna pattern given by
\begin{equation}
F^{2} = \frac{1}{4}(1+cos^{2}(\theta))^{2}cos^{2}(2\phi)+cos^{2}(\theta)sen^{2}(2\phi),
\end{equation}
with the angles $\theta$ and $\phi$ varying in the range $[0,\frac{1}{2}\pi]$ \cite{BSchutz}. Moreover, the Power Spectrum Density is given by
\begin{equation}
    \frac{S_{h}(x)}{S_0}=(4.49x)^{-56} + 0.16x^{-4.52} + 0.52 + 0.32x^2,
\end{equation}
with $S_{0} / Hz^{-1}=9.0 \times 10^{-46}$, $f_{s}/Hz = 40$ and $f_{0}/Hz = 150$ \cite{BSathyaprakash}.

The second interferometer configuration has a triangular pattern (as the Einstein Telescope). Besides the difference in angles, it also has a difference in sensitivity. In this case the antenna pattern is given by
\begin{equation}
F^2 = \frac{9}{256}\left(35+28cos(2\theta)+cos(4\theta)\right),
\end{equation}
with the angles $\theta$ varying between [0,$\pi$] \cite{WZhao}. Its Power Spectrum Density is given by 
\begin{equation}
Sh(f) = S_0 \left[x^{p_1} + a_1x^{p_2} + a_2 \frac{1 + b_1x + b_2x^2 + b_3x^3 + b_4x^4 + b_5x^5 + b_6x^6}{1 + c_1x + c_2x^2 + c_3x^3 + c_4x^4}\right]
\end{equation}
where $x = f/f_0$ with $f_0 = 200Hz$ and $S_0 = 1.449\times10^{-52}Hz^{-1}$. The other parameters are as follows: $p_1 = -4.05$, $p_2 = -0.69$, $a_1 = 185.62$, $a_2 = 232.56$, 
$b_1 = 31.18$, $b_2 = -64.72$, $b_3 = 52.24$, $b_4 = -42.16$, $b_5 = 10.17$, $b_6 = 11.53$, $c_1 = 13.58$, $c_2 = -36.46$, $c_3 = 18.56$, $c_4= 27.43$ \cite{WZhao}.

The amplitude of the wave is needed to calculate the SNR, and it is given by \cite{WZhao}
\begin{equation}
A^2 = \frac{4}{D_{L}^{2}}\left(\frac{5\pi}{96}\right)\pi^{-\frac{7}{3}}\left(M_{c}G(1+z)\right)^\frac{5}{3},
\end{equation}
the $M_c$ stands for the chirp mass of the gravitational wave, which depends on the masses of the two merging objects, whether neutron stars (NS) or black holes (BH). The formula for the chirp mass is given by
\begin{equation}
M_{c}=\mu^{\frac{3}{5}}M^{\frac{2}{5}}=\frac{(m_{1}m_{2})^{\frac{3}{5}}}{(m_{1}+m_{2})^{\frac{1}{5}}}
\end{equation}
we define the components masses of the binary system $m_1$ and $m_2$, then $M = m_1 + m_2$ as the total mass and $\mu = (m_{1}m_{2})/{M}$ as the reduced mass~\cite{MMaggiore}. The mass ranges for each component are $[1,3]M_\odot$ for neutron stars and $[5,35]M_\odot$ for black holes, with both obeying the respective probability density functions \cite{PLandry,WFarr}.

\subsection{Hubble Parameter}

As for our sample of $H(z)$, we adopt a compilation of 30 measurements obtained from differential galaxy ages, and 18 measurements from the radial BAO mode. This data-set is hereafter named CC, as they are commonly referred in the literature as cosmic chronometers. Instead of using the actual observational measurements, we replace them by a realization of the flat $\Lambda$CDM model, i.e., $H^{\rm obs}(z) \rightarrow \mathcal{N}(H^{\rm fid}(z), \sigma_{H^{\rm obs}(z)})$, so that the Hubble parameter follows the Friedmann equation
\begin{equation}\label{eq:hz_fid}
\left[\frac{H^{\rm fid}(z)}{H^{\rm fid}_0}\right]^2 = \Omega^{\rm fid}_{\rm m}(1+z)^3 + \Omega^{\rm fid}_{\Lambda} \,,
\end{equation} 
and again, we assume a fiducial flat $\Lambda$CDM model consistent with the Pantheon+SH0ES best-fit as Section II.B~\cite{PantheonPlus1}.
By the same token, we simulate future $H(z)$ measurements expected from ongoing redshift surveys, like J-PAS. We produce 30 data points from a realization of the $\Lambda$CDM fiducial model, as done for the observational CC data-set, but assuming that the $H(z)$ uncertainties should follow the values shown in Fig. 15 of~\cite{J-PAS2} for a J-PAS 8500 degree$^2$ configuration. Also, we assume that these data points should follow a redshift distribution $P(z)$ in the interval $0.3<z<2.5$, as given by~\cite{Wang20,Bengaly23}
\begin{equation}\label{eq:pz}
P(z; \; k,\theta) = z^{k-1}\frac{e^{-z/\theta}}{\theta^{k}\Gamma(k)} \;,
\end{equation}
where we fix $\theta$ and $k$ to their respective best fits to the real data, i.e., $\theta_{\rm bf}=0.647$ and $k=1.048$.  

\section{Results}

\begin{figure*}[!t]
\centering
\includegraphics[scale=0.4]{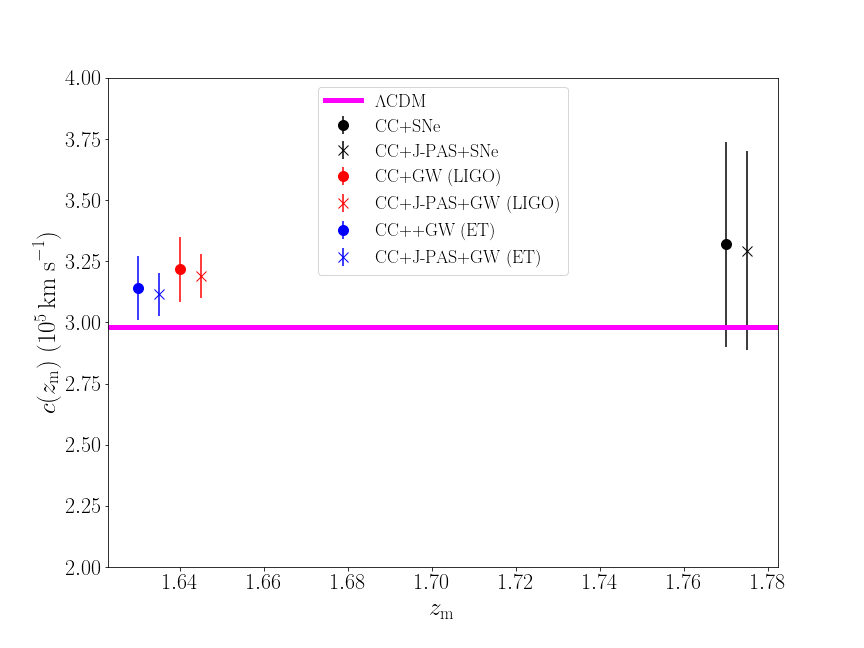}
\includegraphics[scale=0.4]{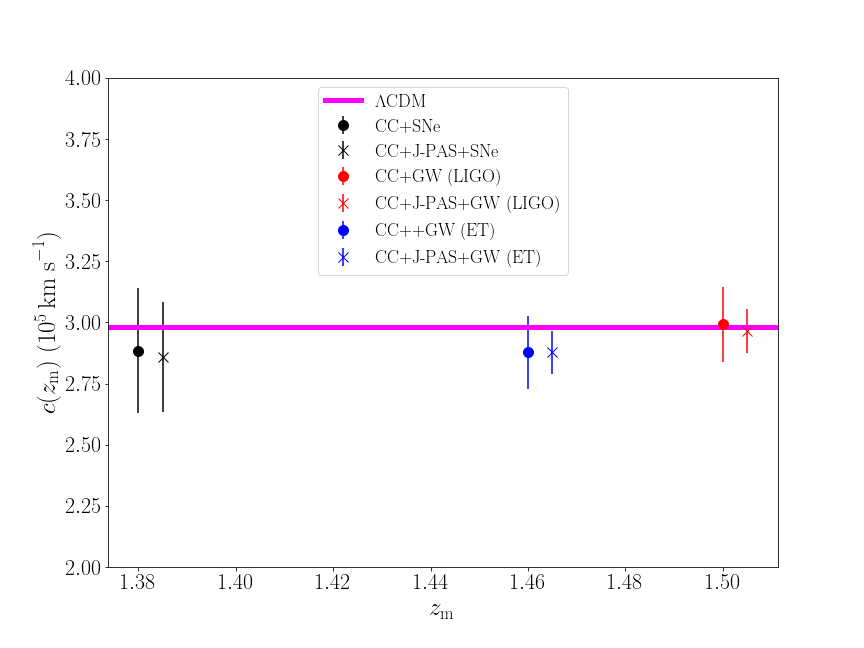}
\caption{{\it Upper panel:} $c(z_{m})$ measurements, as a function of $z_{m}$, obtained assuming the Squared Exponential GP kernel from the following combination of data-sets: CC+SNe (black circle), CC+GW from LIGO (red circle), CC+GW from ET (blue circle). The combination of the same data-sets with the $H(z)$ from J-PAS are denoted by black, red and blue stars, respectively. {\it Lower panel:} Same as left panel, but rather assuming the Mat\'ern(7/2) kernel. The uncertainties in all cases correspond to the $2\sigma$ CL, and the magenta horizontal line represent the value predicted by the $\Lambda$CDM scenario.}
\label{fig:cz_measurements}
\end{figure*}

In order to obtain $c(z_m)$ and its respective uncertainty, as in Eq.~\ref{eq:czm} and~\ref{eq:sigma_czm}, respectively, we need to carry out an interpolation across the redshift range that is covered by those data-sets. So, we follow the approach of~\cite{GC}, and reconstruct $\{D_{\text{A}}(z)$, $H(z)\}$, as well as their respective derivatives $\{D'_{\text{A}}(z)$, $H'(z)\}$, from those datasets using a non-parametric approach -- namely the Gaussian Processes (GP) method, as in the G$\scriptsize \text{A} $PP package \cite{MSeikel}. We assume two kernels to perform these reconstructions, i.e., Squared Exponential (hereafter SqExp) and Mat\'ern(7/2) (hereafter Mat72), for $n = 250$ bins within the redshift range $0.01 < z < 2.5$, in which we optimize the GP hyperparameters in both cases. So we can obtain $z_{m}$ at the point where the reconstructions yield $D'_{A} \simeq 0$, and then calculate $c(z_m)$ and $\sigma_{c(z_m)}$ according to Eqs.~\eqref{eq:czm} and~\eqref{eq:sigma_czm}. More details on the reconstructions obtained for each case are shown in the Appendix.

\begin{table}[!h]
\centering
\caption{Results for the $c(z_{\rm m})$ measurements assuming the Squared Exponential GP kernel. The first column displays the combination of data-sets, the second column shows the reconstructed $z_{\rm m}$ value, the third column provides the $c(z_{\rm m})$ measurements and their uncertainties in $1\sigma$ CL, in units of $10^5 \, \mathrm{km \, s}^{-1}$, and the fourth column gives their relative uncertainty, in percent.}
\vspace{0.3cm}
\begin{tabular}{|c|c|c|c|}
\hline 
data-sets (CC+) & $z_{\rm m}$ & $c(z_{\rm m}) \pm 
\sigma_{c(z_{\rm m})}$ & uncertainty (\%) \\ 
\hline 
+SNe (Pantheon+SH0ES) & $1.77$ & $3.319 \pm 0.210$ & $6.25$ \\ 
+GW (LIGO) & $1.64$ & $3.217 \pm 0.067$ & $2.08$ \\ 
+GW (ET) & $1.63$ & $3.140 \pm 0.066$ & $2.10$ \\ 
\hline 
data-sets (CC+JPAS+) & $z_{\rm m}$ & $c(z_{\rm m}) \pm \sigma_{c(z_{\rm m})}$ & uncertainty (\%) \\ 
\hline 
+SNe (Pantheon+SH0ES) & $1.77$ & $3.293 \pm 0.203$ & $6.16$ \\ 
+GW (LIGO) & $1.64$ & $3.190 \pm 0.044$ & $1.38$ \\ 
+GW (ET) & $1.63$ & $3.114 \pm 0.043$ & $1.38$ \\
\hline
\end{tabular} 
\label{tab:cz_measurements_sqexp_traintrue}
\end{table}

\begin{table}[!h]
\centering
\caption{Same as Table~\ref{tab:cz_measurements_sqexp_traintrue}, but rather assuming the Mat\'ern(7/2) kernel.}
\vspace{0.3cm}
\begin{tabular}{|c|c|c|c|}
\hline 
data-sets (CC+) & $z_{\rm m}$ & $c(z_{\rm m}) \pm 
\sigma_{c(z_{\rm m})}$ & uncertainty (\%) \\ 
\hline 
+SNe (Pantheon+SH0ES) & $1.38$ & $2.884 \pm 0.128$ & $4.45$ \\ 
+GW (LIGO) & $1.50$ & $2.992 \pm 0.077$ & $2.59$ \\ 
+GW (ET) & $1.46$ & $2.877 \pm 0.074$ & $2.58$ \\ 
\hline 
data-sets (CC+JPAS+) & $z_{\rm m}$ & $c(z_{\rm m}) \pm \sigma_{c(z_{\rm m})}$ & uncertainty (\%) \\ 
\hline 
+SNe (Pantheon+SH0ES) & $1.38$ & $2.859 \pm 0.112$ & $3.91$ \\ 
+GW (LIGO) & $1.50$ & $2.965 \pm 0.045$ & $1.52$ \\ 
+GW (ET) & $1.46$ & $2.851 \pm 0.043$ & $1.51$ \\ 
\hline
\end{tabular} 
\label{tab:cz_measurements_mat72_traintrue}
\end{table}

Figure~\ref{fig:cz_measurements} displays all these results in black (CC+SNe), blue (CC+GW from LIGO), and red (CC+GW from ET) circles, whereas the crosses of the same color code represent the combination of the corresponding data-sets with the J-PAS simulations, albeit at a $2\sigma$ CL uncertainty for the sake of enhancing visualization. The difference between the left and the right panel is the GP kernel under assumption -- the left panel shows the results obtained from the SqExp kernel reconstructions, while the right panel displays the Mat72 case. The horizontal line represents the locally measured $c$, i.e., $c = 2.998 \times 10^5 \, \mathrm{km \, s}^{-1}$. So we can clearly see that our results are in agreement with this local value, and they are consistent with each other, with the results obtained from Mat72 being slightly more compatible with the local measurements. 

In tables~\ref{tab:cz_measurements_sqexp_traintrue} and~\ref{tab:cz_measurements_mat72_traintrue}, we present the results obtained from the GP reconstructions assuming, respectively, the SqExp and Mat72 kernels. We can see in the top of both tables that the combination of current CC+SNe data provides a measurement of $c(z_{\text{m}})$ with a precision of $\sim 6.2\%$ (SqExp), and $\sim 4.5\%$ (Mat72), whereas combining CC with GW simulations from ET and LIGO, the precision is slightly improved -- $\sim 2\%$ (SqExp), and $\sim 2.5\%$ (Mat72). Also, we note that the results obtained by LIGO and ET are totally consistent with each other, to a sub-percent level. When we include the simulated J-PAS $H(z)$ measurements to the CC data-set, and combine it with the SN and GW data, as shown in the bottom of both tables, we find that the precision of the $c(z_{\text{m}})$ measurement is only mildly improved in the former case, but significantly improved for the latter, reaching $\sim 1.4\%$ (SqExp), and $\sim 1.5\%$ (Mat72). 

These results demonstrate that future observational data have the capability of improving the measurements of the speed of light to a few percent level, and therefore we will be able determine with much higher precision whether its value actually agrees with the measurements carried out in local laboratories, as predicted from the standard cosmological model, or whether there is any hint at new physics if found otherwise.

\section{Conclusions}

Performing cosmological measurements of the fundamental physical constants is of great importance, as any statistically significant departure from local measurements would immediately require a reformulation of the standard cosmological model. In this work, we measure the speed of light $c$ with Hubble parameter and angular diameter distance measurements from current data-sets, as obtained from a compilation of galaxy ages and radial baryon acoustic oscillations for the former, and Type Ia Supernova distances from Pantheon+SH0ES for the latter. We do so by performing a Gaussian Process reconstruction of such quantities, in order to avoid the assumption of a cosmological model. Then, we forecast the precision of such a measurement by simulating the same kind of data from upcoming galaxy redshift surveys, such as J-PAS, and from standard sirens from gravitational wave experiments, as in the cases of LIGO and the Einstein Telescope.

We obtained a significant improvement between the current and future constraints, reducing our uncertainties from $4-6\%$ to about $2-2.5\%$ when the gravitational wave simulations are considered, and to nearly $1.5\%$ when the simulated J-PAS measurements are taken into account as well, at redshifts around $1.4 < z < 1.8$. We noted that these figures hold regardless of the assumption on the reconstruction kernel, the number of bins for each reconstruction etc.

Our result show the capability of the upcoming generation of cosmological observations in terms of carrying out a powerful test of fundamental physics, as we found that such a combination of data-sets, along with the statistical approach herein deployed, will be able to provide a percent-level precision measurement of the speed of light at cosmological scales. Therefore, we will be able to verify if the speed of light at high redshift ranges -- which corresponds to a Universe age of around $3.2 - 4.0$ Gyr -- is actually consistent with local measurements with significantly improved precision, which will help determining the validity (or departure) of the standard model of Cosmology.

{\it Acknowledgments:} 
We thank to Jailson S. Alcaniz for valuable discussions. JS acknowleges financial support from Coordenação de Aperfeiçoamento de Pessoal de Nível Superior (CAPES). CB acknowledges financial support from Funda\c{c}\~ao \`a Pesquisa do Estado do Rio de Janeiro (FAPERJ) - Postdoc Recente Nota 10 (PDR10) fellowship. JM acknowledges financial support from Conselho Nacional de Desenvolvimento Científico e Tecnológico (CNPQ) - Undergraduate research fellowship. RSG thanks financial support from the Funda\c{c}\~ao de Amparo \`a Pesquisa do Estado do Rio de Janeiro (FAPERJ) grant SEI-260003/005977/2024 - APQ1.



\begin{appendix}

\section{Gaussian Process Reconstructions}
\label{sec:gp_rec}

\begin{figure*}[]
\centering
\includegraphics[width=0.49\textwidth, height=7.0cm]{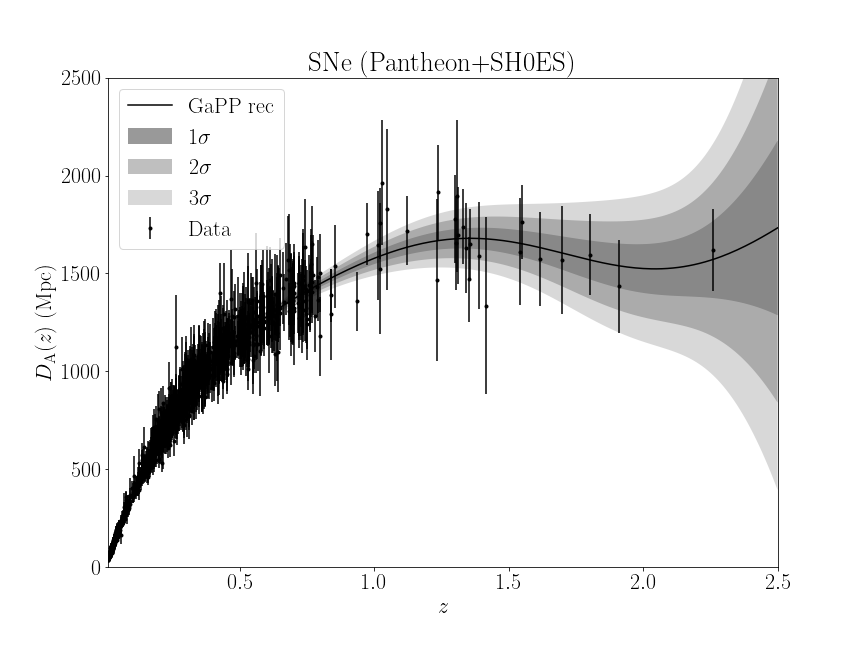}
\includegraphics[width=0.49\textwidth, height=7.0cm]{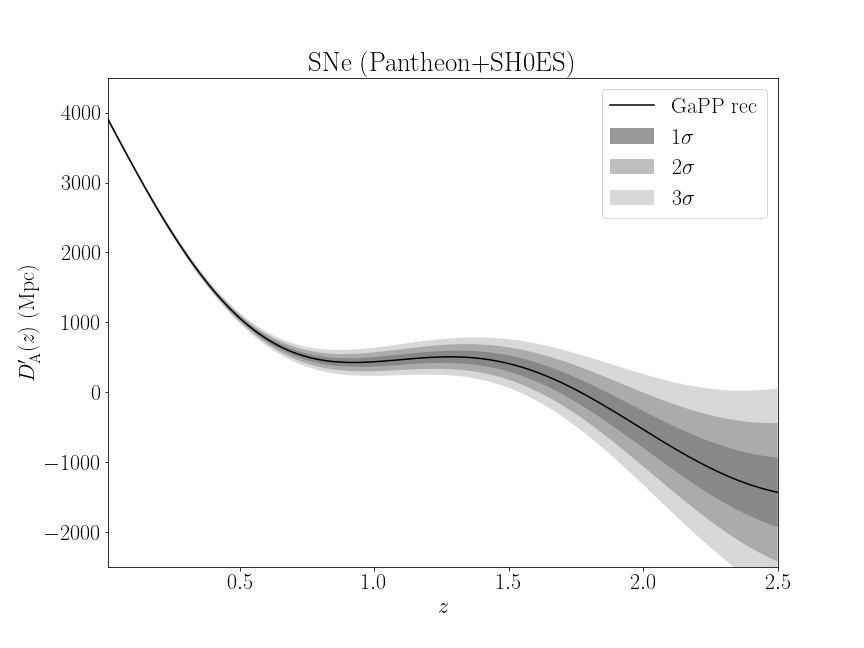}
\includegraphics[width=0.49\textwidth, height=7.0cm]{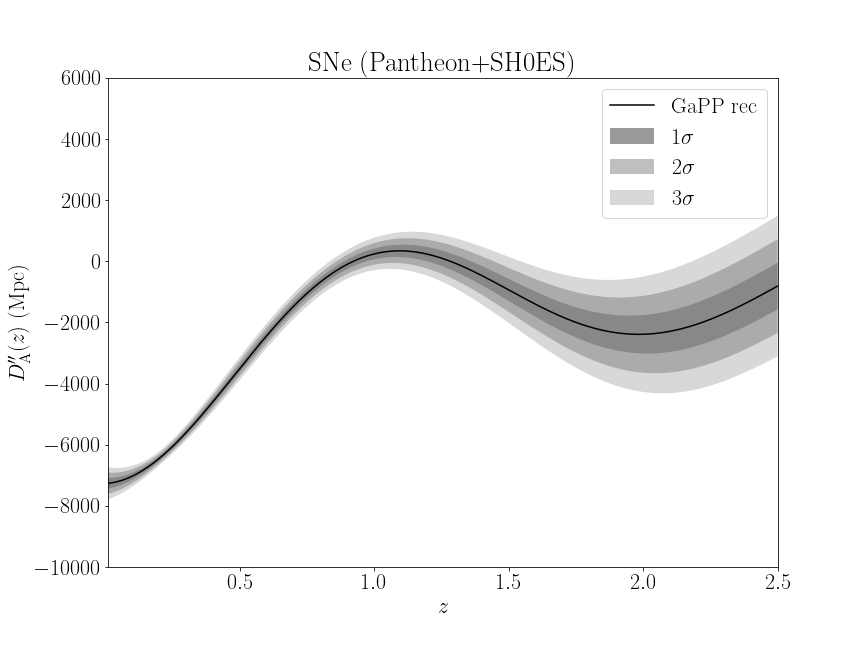}
\caption{The top left panel shows the Gaussian Process reconstruction of the angular diameter distance, $D_A(z)$, as obtained by Eqs.~\eqref{eq:MB} to~\eqref{eq:DAz}, using the Pantheon+SH0ES Type Ia Supernova compilation. The top right panel displays the reconstruction of the first order derivative, $D_A'(z)$, whereas the bottom central panel depicts the reconstruction of the second order derivative, $D_A''(z)$, still obtained from the same observational sample. The black dots with error bars in the $D_A(z)$ plot represent the observational data, and the gray curves denote the $1$, $2$, and $3\sigma$ confidence level of the reconstructions, respectively from the darker to the lighter shade. All reconstructions were performed assuming the Squared Exponential kernel.}
\label{fig:rec_sn_sqexp}
\end{figure*}

\begin{figure*}[]
\centering
\includegraphics[width=0.49\textwidth, height=7.0cm]{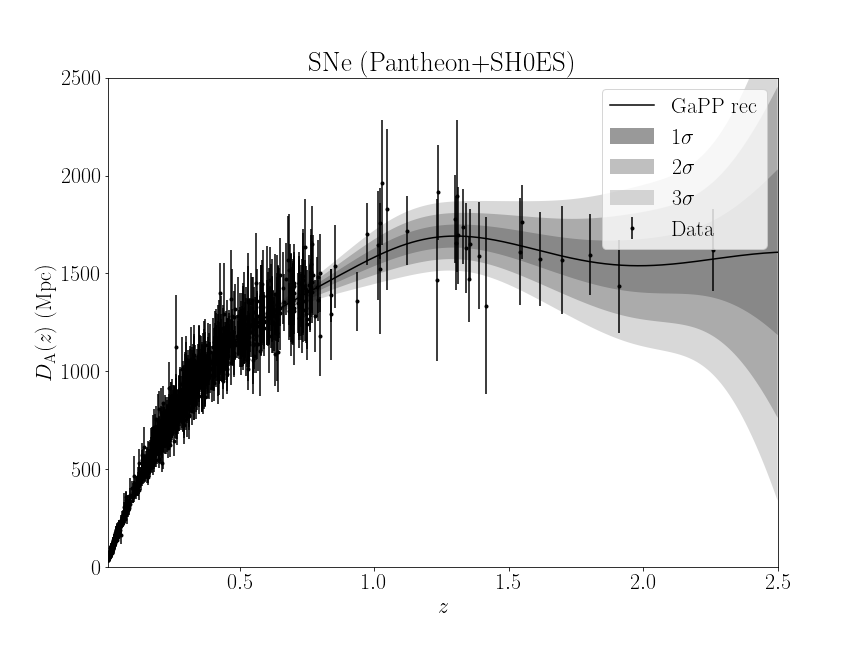}
\includegraphics[width=0.49\textwidth, height=7.0cm]{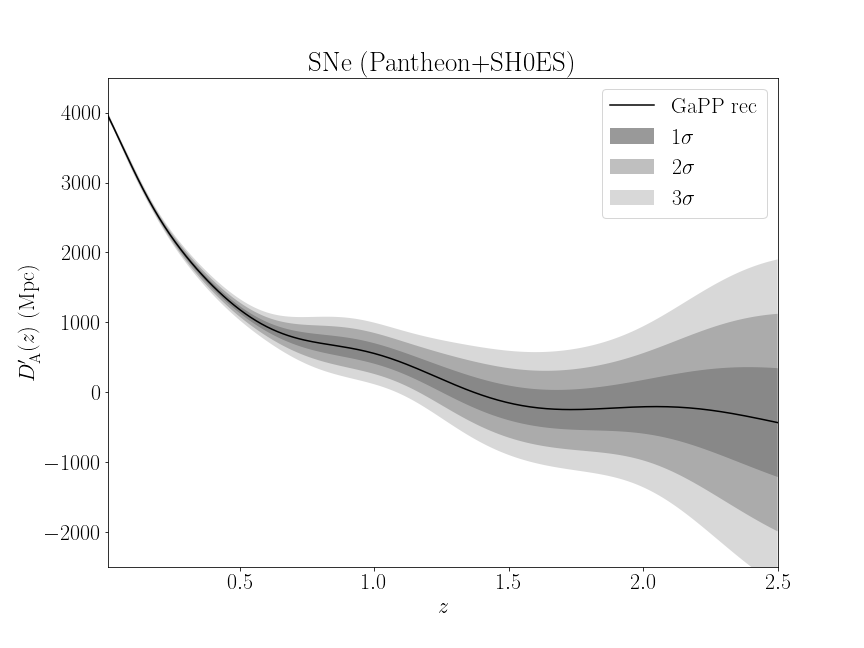}
\includegraphics[width=0.49\textwidth, height=7.0cm]{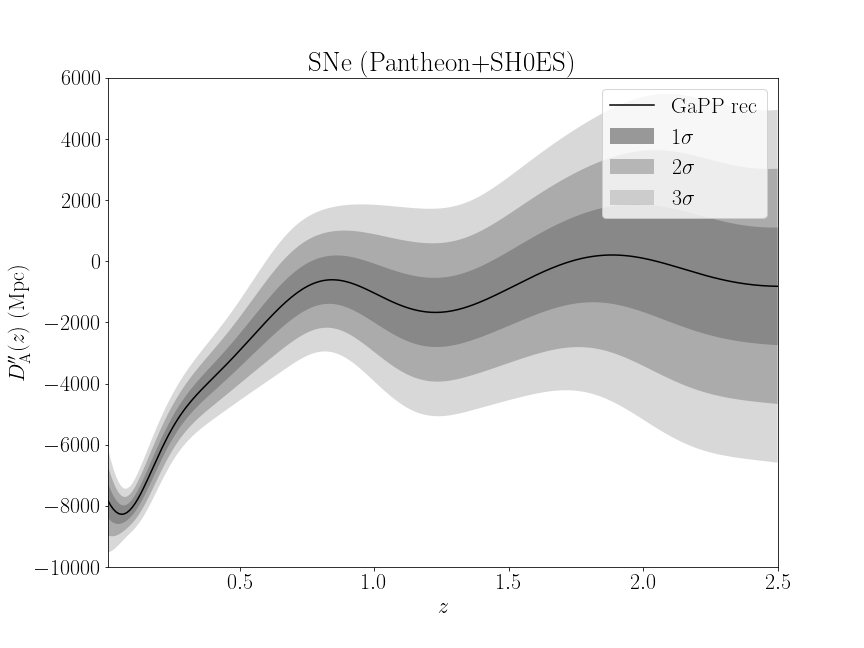}
\caption{Same as Fig.~\ref{fig:rec_sn_sqexp}, but assuming the Mat\'ern(7/2) Gaussian Process kernel instead.}
\label{fig:rec_sn_mat72}
\end{figure*}

\begin{figure*}[]
\centering
\includegraphics[width=0.49\textwidth, height=7.0cm]{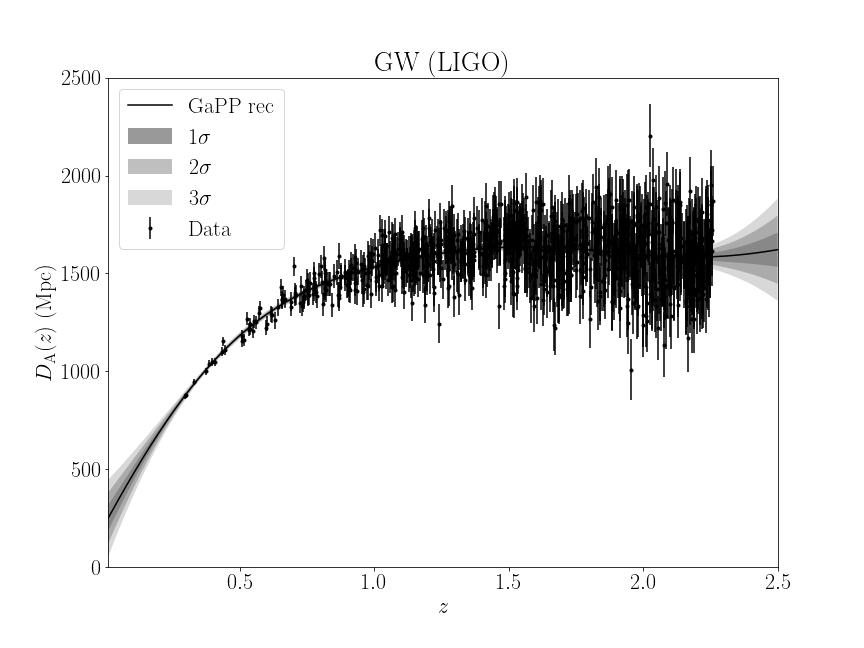}
\includegraphics[width=0.49\textwidth, height=7.0cm]{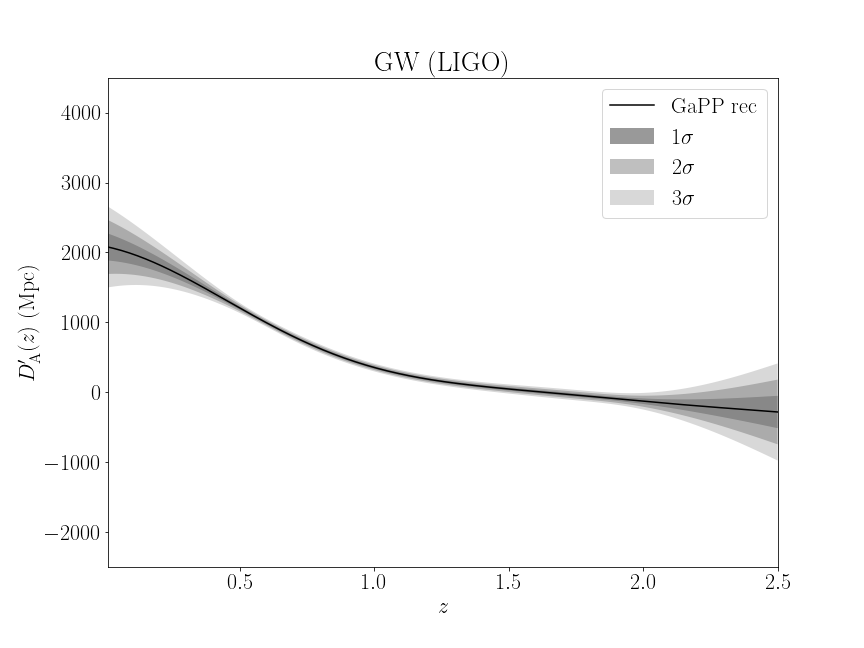}
\includegraphics[width=0.49\textwidth, height=7.0cm]{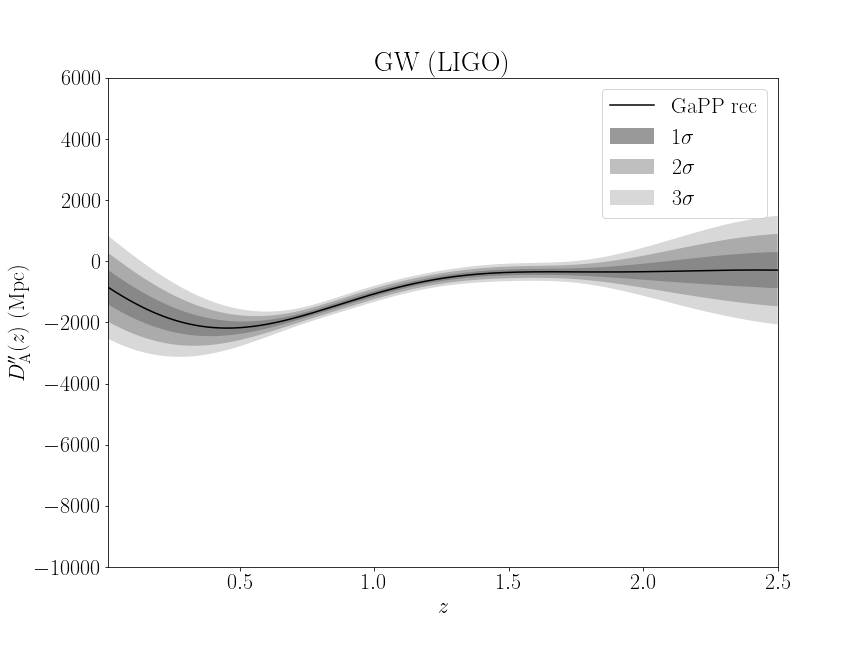}
\caption{Same as Fig.~\ref{fig:rec_sn_sqexp}, but valid for the simulated gravitational wave data-set assuming the LIGO interferometer specifications.}
\label{fig:rec_gwligo_sqexp}
\end{figure*}

\begin{figure*}[]
\centering
\includegraphics[width=0.49\textwidth, height=7.0cm]{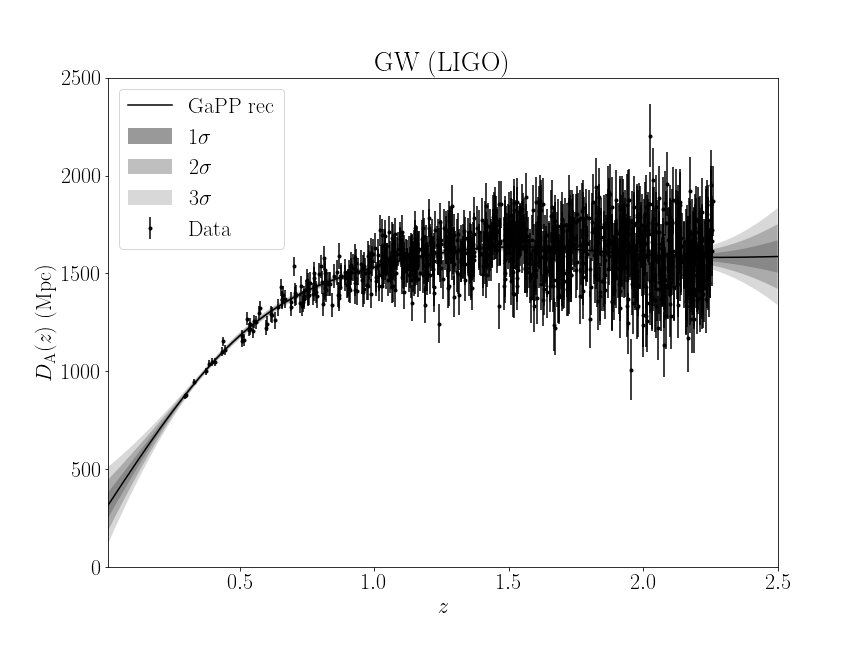}
\includegraphics[width=0.49\textwidth, height=7.0cm]{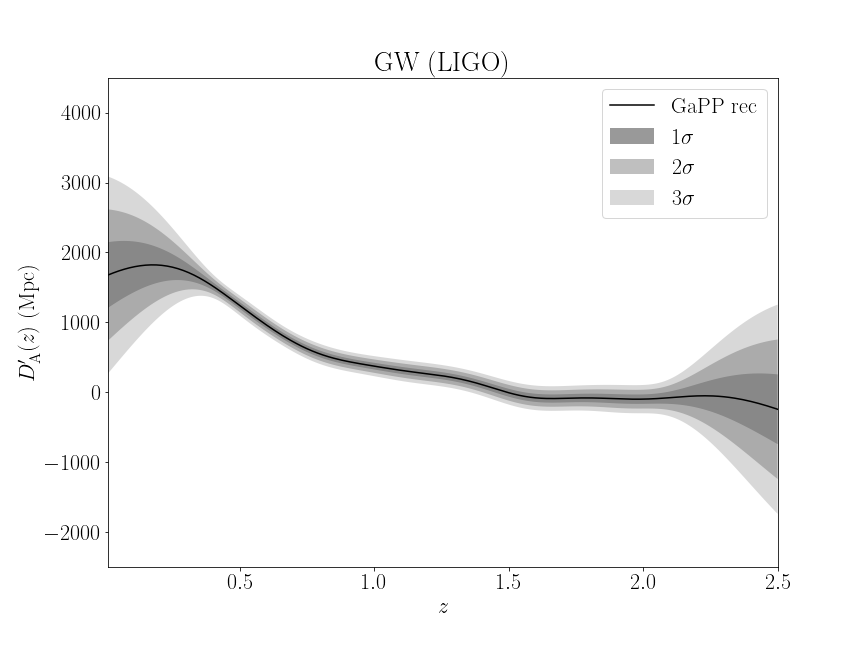}
\includegraphics[width=0.49\textwidth, height=7.0cm]{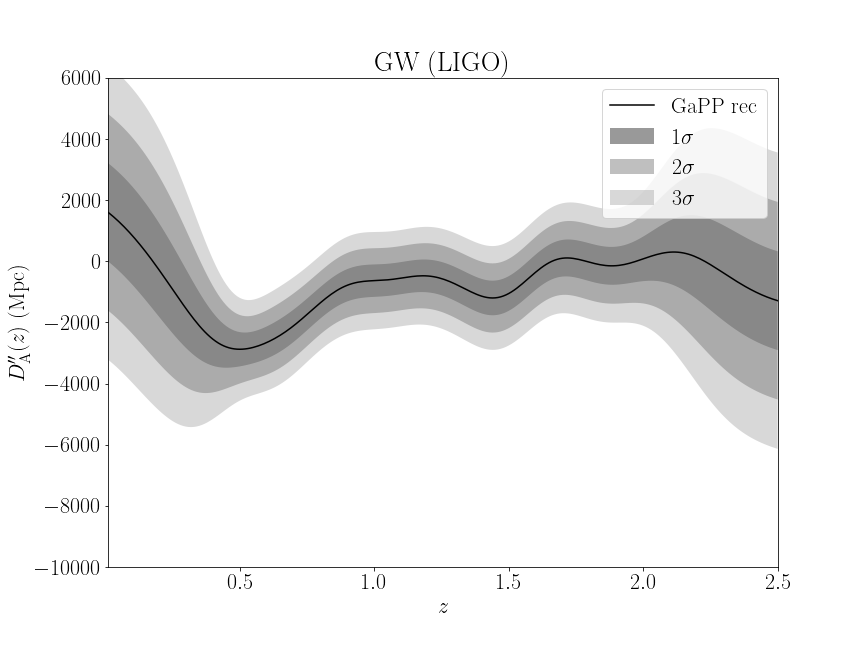}
\caption{Same as Fig.~\ref{fig:rec_gwligo_sqexp}, but assuming the Mat\'ern(7/2) Gaussian Process kernel instead.}
\label{fig:rec_gwligo_mat72}
\end{figure*}

\begin{figure*}[]
\centering
\includegraphics[width=0.49\textwidth, height=7.0cm]{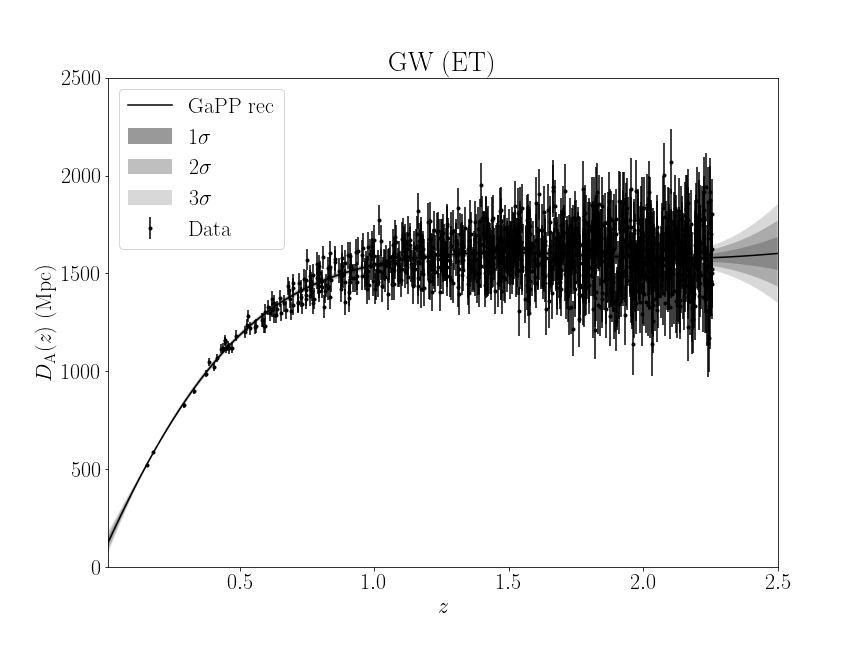}
\includegraphics[width=0.49\textwidth, height=7.0cm]{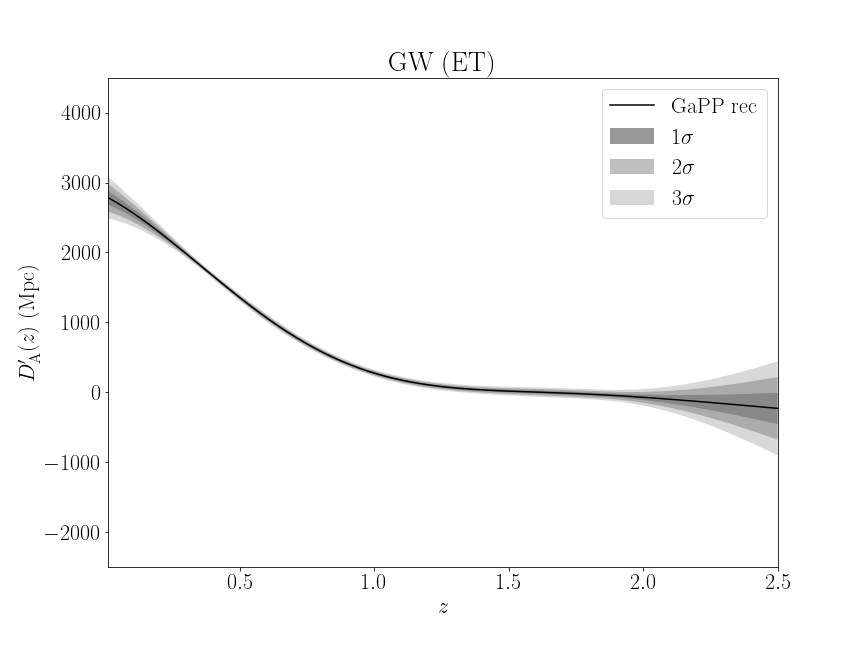}
\includegraphics[width=0.49\textwidth, height=7.0cm]{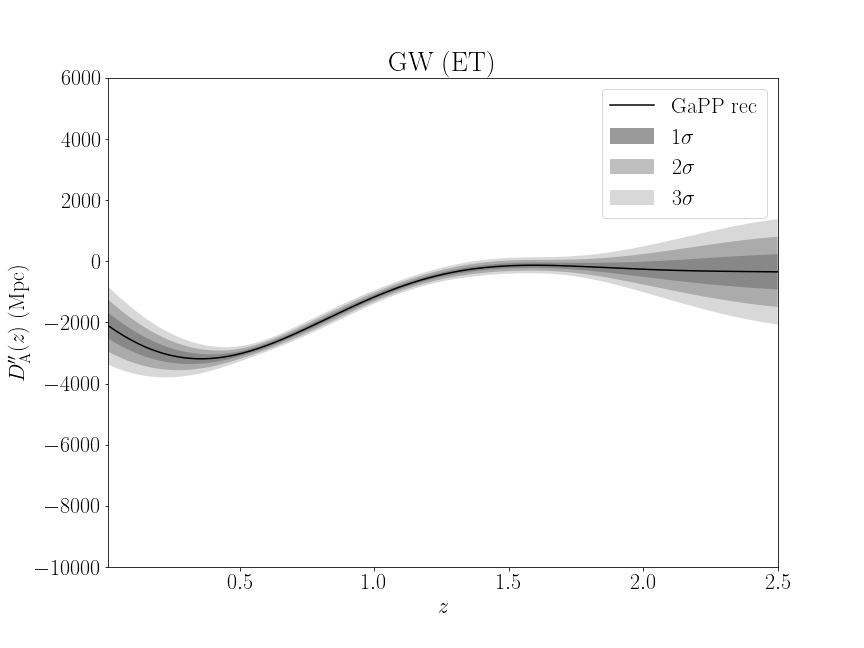}
\caption{Same as Fig.~\ref{fig:rec_gwligo_sqexp}, but valid for the simulated gravitational wave data-set assuming the Einstein Telescope specifications.}
\label{fig:rec_gwet_sqexp}
\end{figure*}

\begin{figure*}[]
\centering
\includegraphics[width=0.49\textwidth, height=7.0cm]{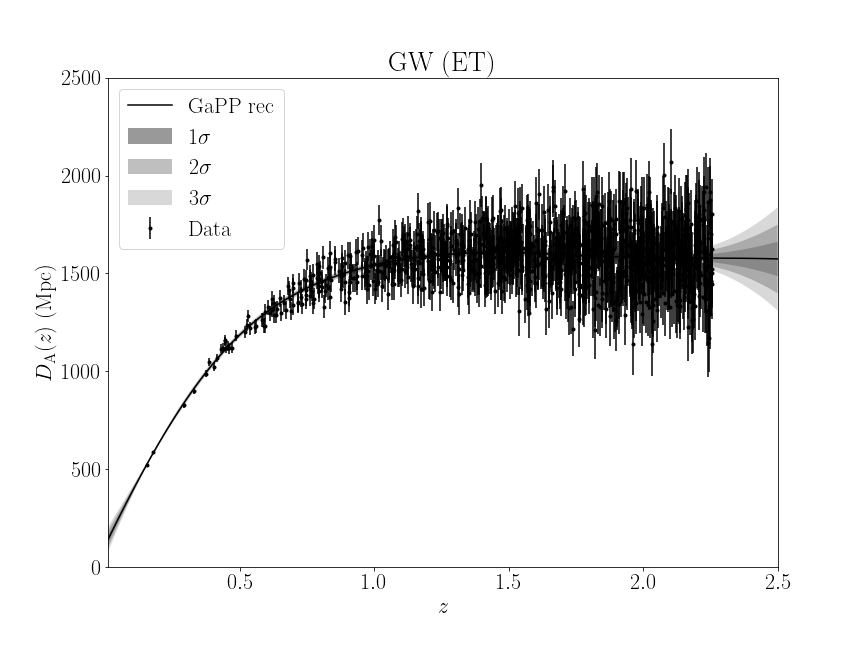}
\includegraphics[width=0.49\textwidth, height=7.0cm]{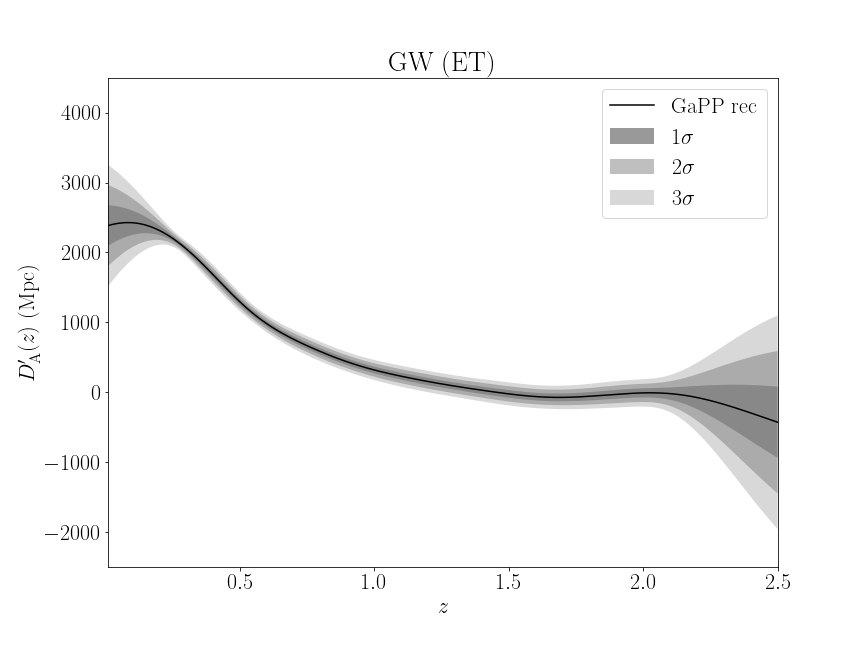}
\includegraphics[width=0.49\textwidth, height=7.0cm]{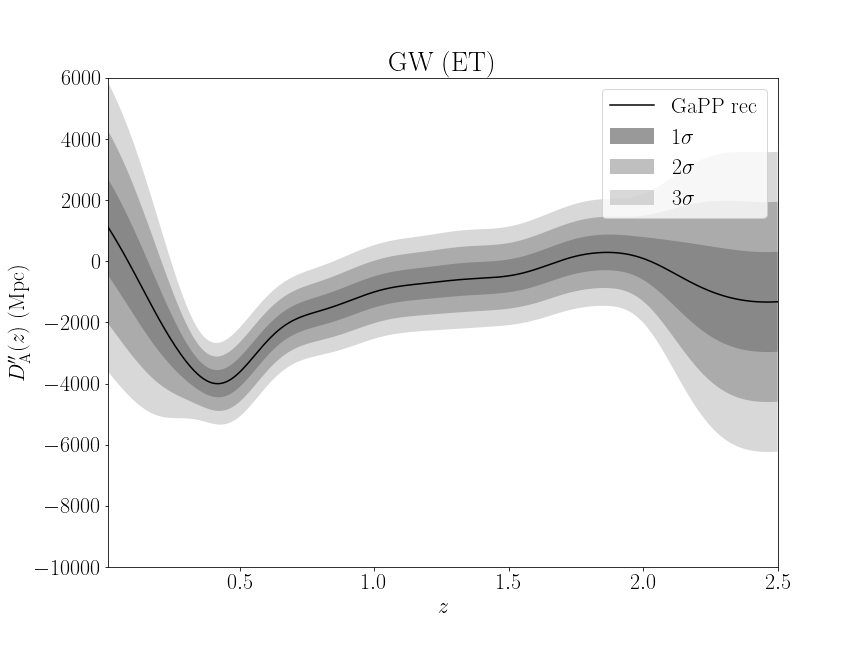}
\caption{Same as Fig.~\ref{fig:rec_gwet_sqexp}, but assuming the Mat\'ern(7/2) Gaussian Process kernel instead.}
\label{fig:rec_gwet_mat72}
\end{figure*}

\begin{figure*}[]
\centering
\includegraphics[width=0.49\textwidth, height=7.0cm]{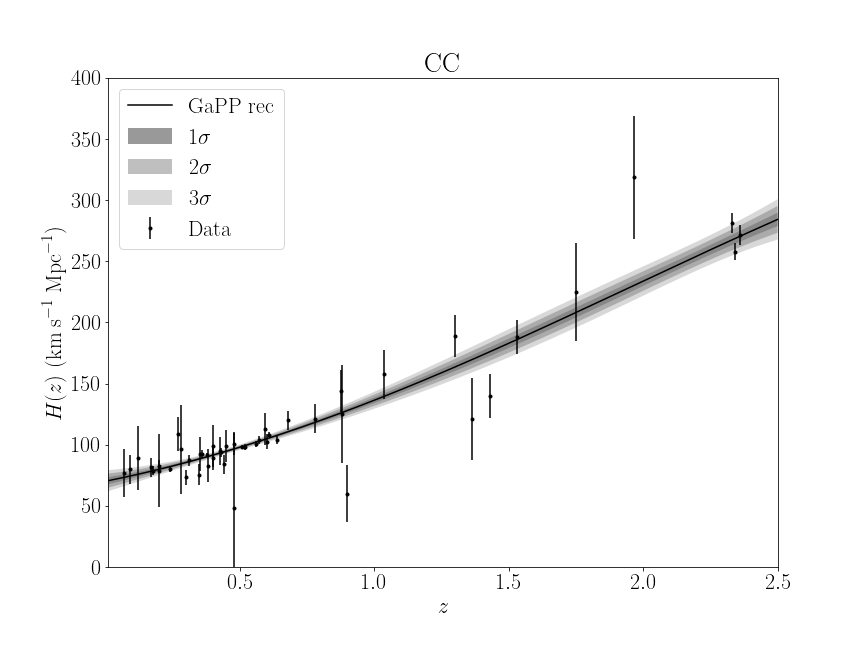}
\includegraphics[width=0.49\textwidth, height=7.0cm]{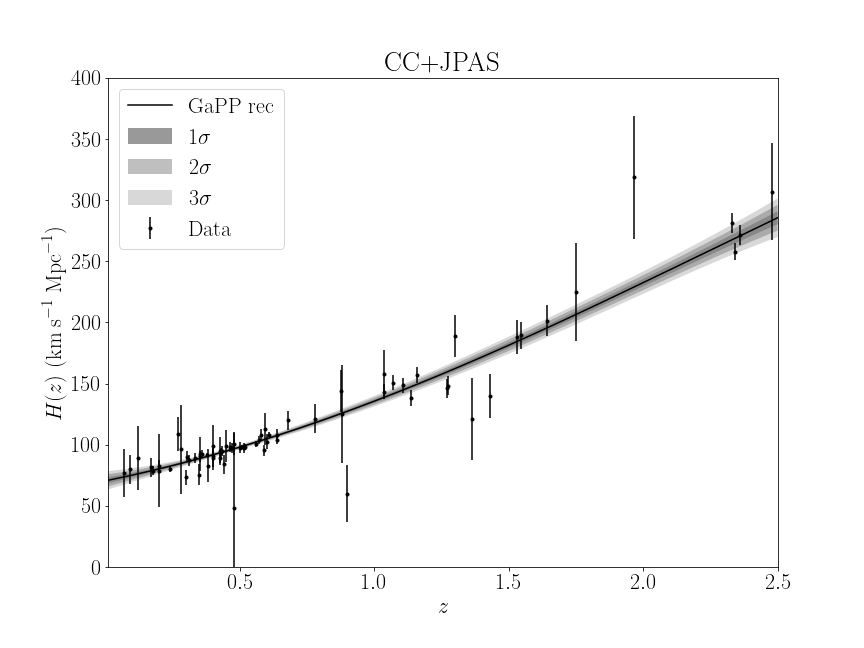}
\includegraphics[width=0.49\textwidth, height=7.0cm]{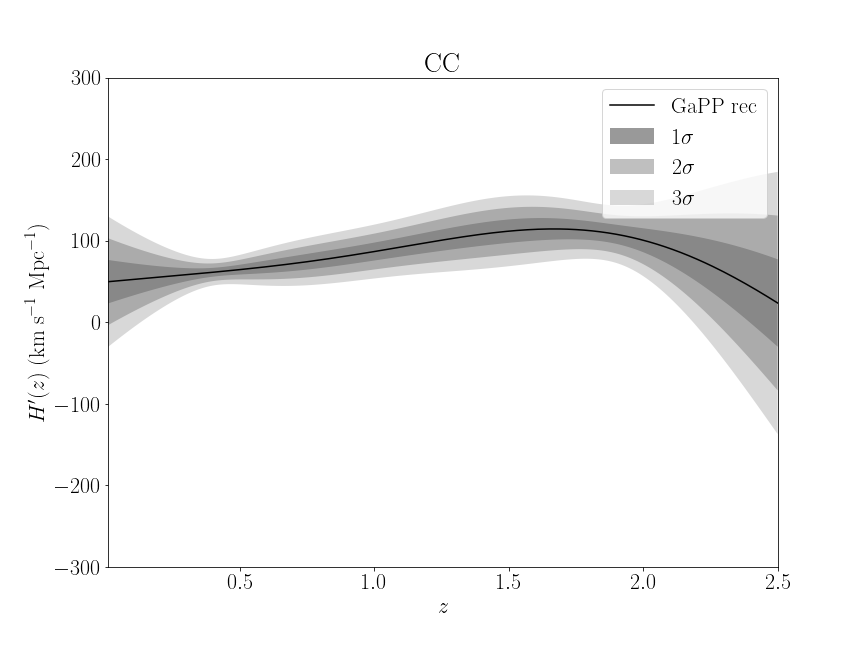}
\includegraphics[width=0.49\textwidth, height=7.0cm]{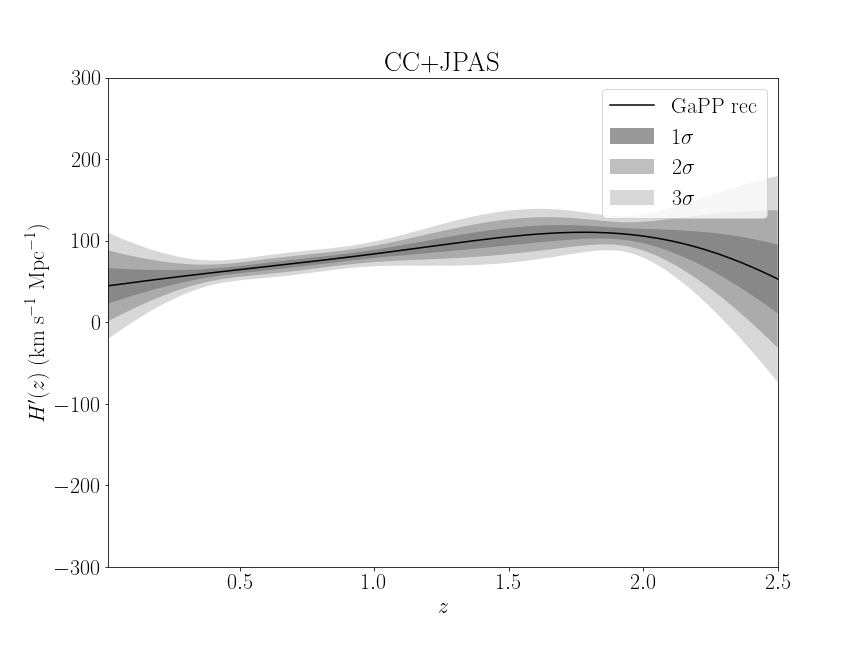}
\caption{{\it Top left panel}: The reconstruction of the Hubble parameter, $H(z)$, obtained from the cosmic chronometer data-set. The black dots depict the data points with their respective uncertainties. {\it Top right panel}: Similar to the former, but rather for the cosmic chronometer combined with simulated J-PAS simulated data-set. {\it Bottom left panel}: The first derivative of the Hubble parameter, $H'(z)$, from the cosmic chronometer sample alone. {\it Bottom right panel}: Similar to the previous plot, but for the cosmic chronometer with J-PAS combination. As in the previous plots, the gray curves denote the $1$, $2$, and $3\sigma$ confidence level of the reconstructions, respectively from the darker to the lighter shade. All results were obtained assuming the Squared Exponential kernel.}
\label{fig:rec_hz_sqexp}
\end{figure*}

\begin{figure*}[]
\centering
\includegraphics[width=0.49\textwidth, height=7.0cm]{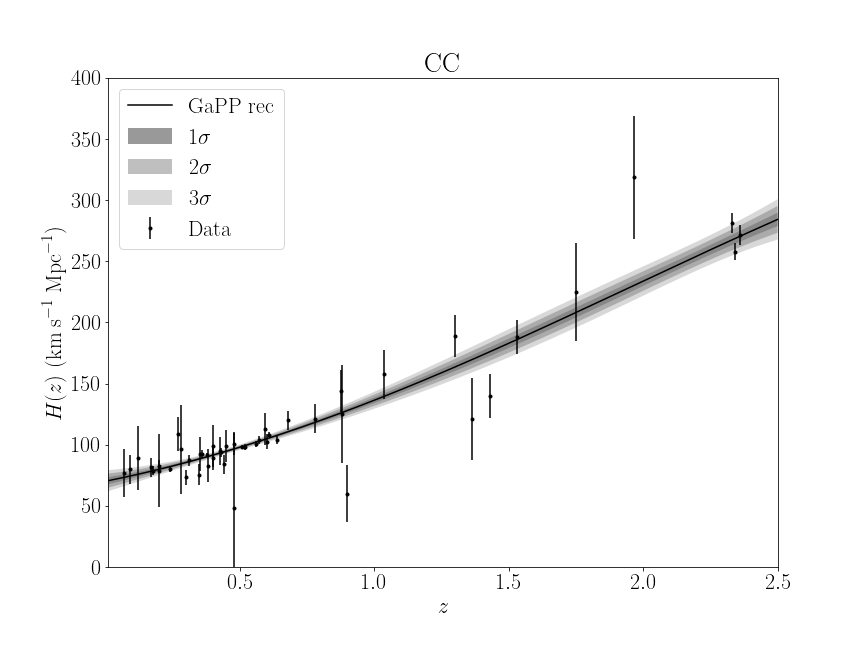}
\includegraphics[width=0.49\textwidth, height=7.0cm]{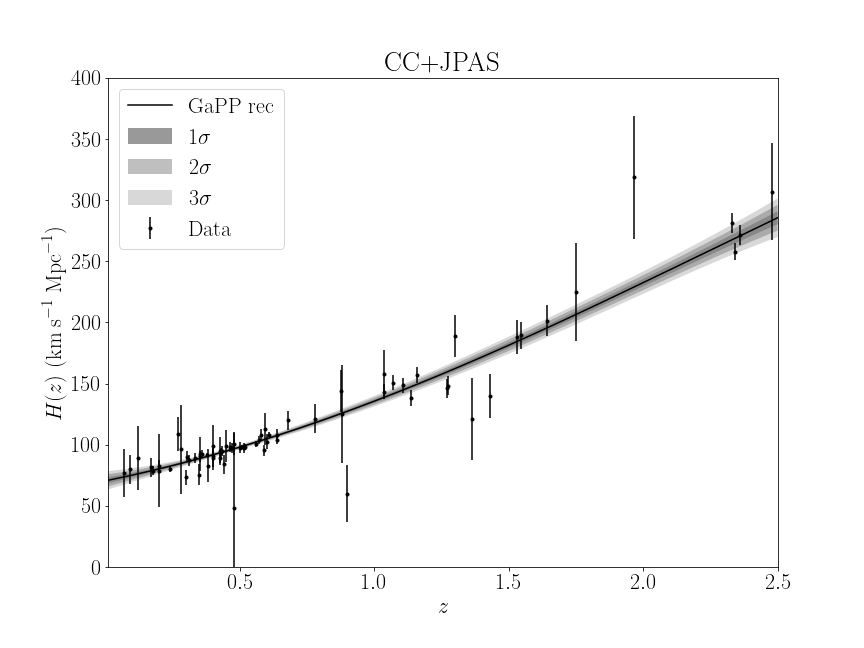}
\includegraphics[width=0.49\textwidth, height=7.0cm]{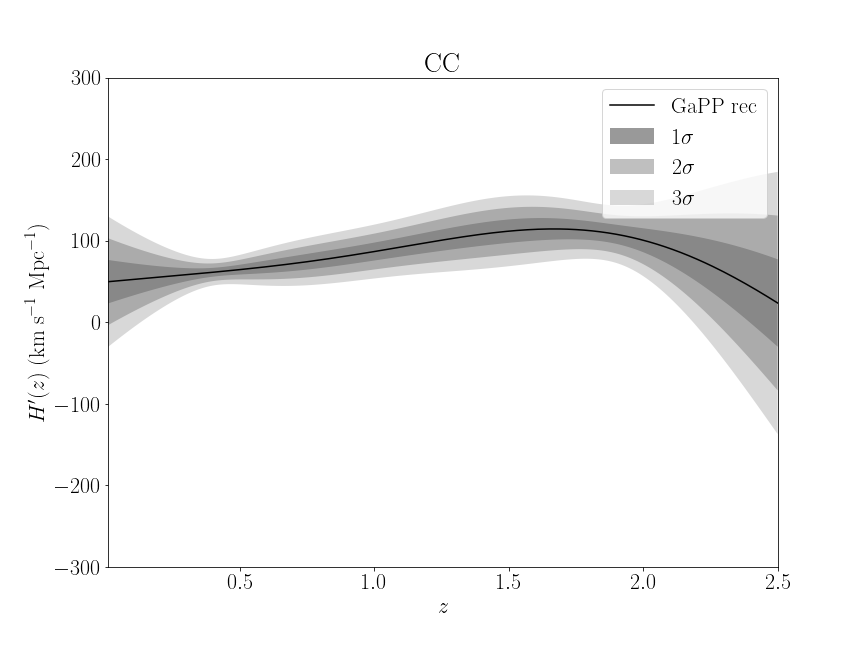}
\includegraphics[width=0.49\textwidth, height=7.0cm]{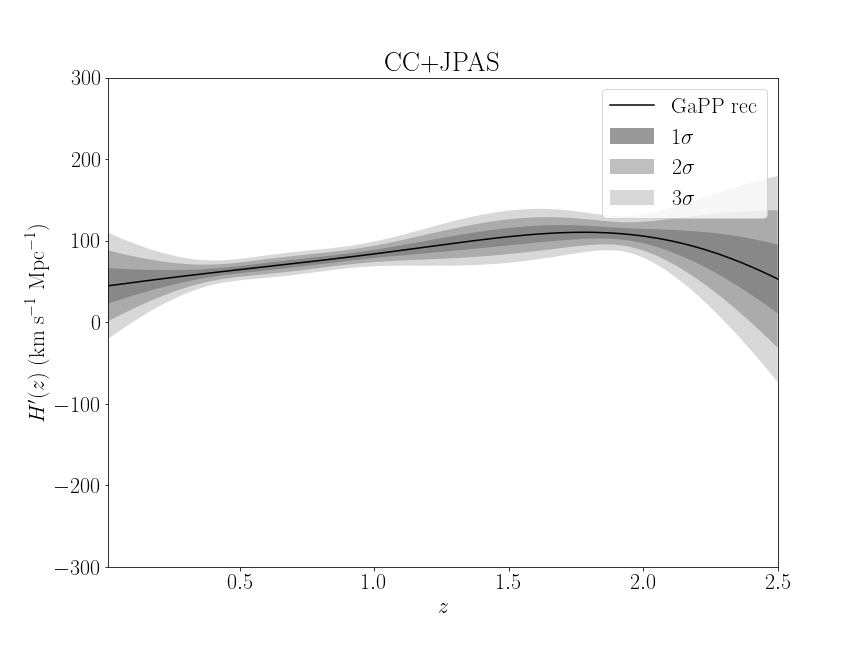}
\caption{Similar to Fig.~\ref{fig:rec_hz_sqexp}, but assuming the Mat\'ern(7/2) Gaussian Process kernel instead.}
\label{fig:rec_hz_mat72}
\end{figure*}

\end{appendix}

\end{document}